\documentstyle[epsf]{fbssuppl}
\addtocounter{page}{104}
\SetFinalPage{114}
\def\beq{\begin{equation}}
\def\eeq{\end{equation}}
\def\bea{\begin{eqnarray}}
\def\eea{\end{eqnarray}}

\def\eqlab#1{\label{eq:#1}}
\def\figlab#1{\label{fig:#1}}

\def\eref#1{(\ref{eq:#1})}
\def\eqref#1{eq.~(\ref{eq:#1})}
\def\Eqref#1{Eq.~(\ref{eq:#1})}

\def\Figref#1{Fig.~\ref{fig:#1}}

\def\sla#1{#1 \hspace{-2.4mm} \slash}

\def\half{\mbox{\small{$\frac{1}{2}$}}}

\def\barr{\left(\begin{array}{c}}
\def\earr{\end{array}\right)}
\def\bmat{\left(\begin{array}{cc}}
\def\emat{\end{array}\right)}

\def\ga{\gamma} 
\def\de{\delta} \def\De{{\it\Delta}}
\def\veps{\varepsilon}  \def\eps{\epsilon}

 \def\La{{\it\Lambda}}

 \def\Si{{\it\Sigma}}

\def\dd{{\rm d}}
\def\pa{\partial}

\def\ie{{i.e.,}}
\def\eg{{e.g.\ }}

\def\pa{\partial}

\def\no{\nonumber}
\def\BK#1#2{{\it #1}, #2}         
\def\CF#1#2#3#4{#1 {\bf #2}, #4 (#3)}  
\def\ibid {{\it ibid.}}

\def\fbs {Few-Body~Systems}

\def\np {Nucl.~Phys.}
\def\prev {Phys.~Rev.}
\def\prc {Phys.~Rev.~C}
\def\prd {Phys.~Rev.~D}
\def\plett {Phys.~Lett.}

\def\rg{{\rm g}}

 \def\tp{\tilde{p}}
\def\tq{\tilde{q}}

\def\pn{$\pi N$ }

\def\3d{3D}
\title{Lorentz covariance of three-dimensional equations}
\author{V.~Pascalutsa and J.~A.~Tjon}
\institute{Institute for Theoretical Physics, University of Utrecht,
Princetonplein 5, \\
3584 CC Utrecht, The Netherlands}

\begin{document}
\maketitle

\begin{abstract}
We show how the invariance under the charge conjugation and CPT symmetry,
present in the Bethe-Salpeter equation, is lost
in the reduction to certain relativistic
three-dimensional equations. This in particular leads to the
breakdown of the standard Lorentz structure and renormalization
procedures for the resulting single-particle propagators. 
We formulate the equal-time approximation of the Bethe-Salpeter equation
in the form which manifestly satisfies the above symmetries, and
apply it to the description of the pion-nucleon interaction 
in a dynamical model based on hadron exchanges. 
We also consider the one-body limit of various three-dimensional
equations for the case of $t$- and $u$-channel 
one-particle-exchange potential.
\end{abstract}

\section{Introduction}
Quantum field theory (QFT) provides us with a suitable
framework unifying principles of relativistic covariance 
and quantum mechanics. Unfortunately, any 
systematic calculation beyond perturbation theory is extremely
complicated within QFT. 
On the other hand, in the ordinary quantum 
mechanics the scattering and bound state problems are well understood 
in terms of the Schr\"odinger or Lippmann-Schwinger  equation.
Its relativistic generalizations, referred to
as {\it quasipotential} (QP) approximations to field theory, 
can therefore be very useful for
practical studies of relativistic effects in the strongly interacting
systems.

The QP equations can conveniently be obtained from the manifestly covariant 
four-dimensional Bethe-Salpeter (BS) equation by approximating the kernel
in some way.
This approximation involves an assumption about
the singularities of the BS kernel,
after which the integration over the 0-th component (time or energy) 
can easily be done explicitly, leading to the three-dimensional (3-D) equation.
One of the first such reductions of the BS equation was studied by
Salpeter \cite{Sal52}, using the {\it instantaneous} approximation.

Since the covariant reductions can be done in infinitely many 
different ways it is desirable to establish certain 
criteria which would constrain the choice. For instance,
an important property one would like to have for a 
relativistic two-body equation
is the {\it correct one-body limit}, which means that in the limit
when one of the particles becomes infinitely heavy the two-body equation
must reduce to the corresponding equation of motion of the light particle 
(e.g., the Klein-Gordon equation) in an external potential. The fact
that the ladder BS equation does not have the correct one-body limit,
created further motivation for the QP 
approach since the one-body limit can
relatively simply be incorporated into a 3-D equation for the
one-particle-exchange (OPE) potential. Some of the first equations 
of this type were suggested
by Gross \cite{Grs69} and by Todorov \cite{Tod73}. Since then
the one-body limit is regarded as an important criterion,
even though not a very restrictive one, 
as many of the equations can be adjusted to satisfy it.
 In particular, Mandelzweig and Wallace \cite{MaW87} 
incorporated the one-body limit in Salpeter's instantaneous 
equation.

In this contribution we would like to demonstrate the importance 
of the constraint put by {\it charge conjugation} symmetry.
Also, the one-body limit constraint will be examined
at the one loop level for various OPE potentials.
As an application, results of a quasipotential modelling of the pion-nucleon
system will be presented as well.

\section{The role of charge conjugation symmetry}
To begin with, let us recall several basic
definitions concerning the Lorentz group. A general
Lorentz transformation $L$ of a four-momentum is
given by real (pseudo-)orthogonal tensor $\La_{\mu\nu}$, 
and may belong to one of
the following four domains:
\bea
L_+^\uparrow &:& \det\La=+1,\,\,\, \La_{00} \geq +1, \no\\
L_-^\uparrow &:& \det\La=-1,\,\,\, \La_{00} \geq +1, \no\\
L_+^\downarrow &:& \det\La=+1,\,\,\, \La_{00} \leq -1, \\
L_-^\downarrow &:& \det\La=-1, \,\,\,\La_{00} \leq -1. \no
\eea
From these, only  transformations $L_+^\uparrow$ form a group
by themselves, called the proper orthochronous
Lorentz group. The other domains
do not contain the unity element, however their multiplication
with $L_+^\uparrow$ may form a group. 
Fields and corresponding Green functions are transformed
according to unitary representations of the Lorentz group. 
Let us remark, that the full Lorentz group transformations
[which include the proper continuous transformations, 
and the (anti-)unitary  transformations of parity and 
time reversal] do not connect
Green functions defined for positive energy (upper light-cone)
with those for negative energy (lower light-cone). In other words,
applying such transformations to the momentum-space 
Green functions can induce only 
$L_+^\uparrow$ and $L_-^\uparrow$ transformations of 
the relevant four-momenta. To be able to induce 
$L_+^\downarrow$ and $L_-^\downarrow$ transformations of the
four-momenta, one needs to include charge conjugation in addition 
to the above mentioned Lorentz transformations. 

In considering some existing relativistic 3-D equations,
we find that they in general do not yield the correct Lorentz structure. 
For example, the calculated self-energy of a
spin-1/2 particle {\em does not} have the following form,
\beq
\eqlab{form}
\Si (\sla{P})= \sla{P} A(P^2) + B(P^2),
\eeq
where $A$ and $B$ are scalar functions of the invariant $P^2$ only.
At first this is surprising, naively we would expect 
form \eref{form} to come out in any covariant formalism. However,
\Eqref{form} holds only if there is a symmetry under all Lorentz
transformations of the four-momenta. Therefore, in order
to obtain the self-energy consistent with  \Eqref{form},
the relativistic equation in question should be covariant under the 
Lorentz group and charge conjugation.

The four-dimensional BS equation, of course, 
preserves the standard structure, such as \Eqref{form}.
The symmetry can obviously be lost in doing the QP reduction. 
To illustrate this, consider the example of a 
scalar self-energy, given by
\beq
\eqlab{sif}
\Si(P^2)=
i\int\! \frac{\dd^4{q}}{(2\pi)^4}\,
\frac{\Phi (q^2,P^2,P\cdot q)}{[(\half P-q)^2 - m^2+i\veps]\,
[(\half P+q)^2 - m^2+i\veps]},
\eeq
where $P$ is the relevant four-vector, $\Phi$ is an ``interaction function''
which corresponds to the product of the two vertex functions, and which
may also have some particle propagation poles.

We can immediately see that \Eqref{sif} is a function of $P^2$ only:
a sign change of $P$ can be absorbed by a change of the
loop variable $q$ to $-q$. In a QP 
description this
substitution in general cannot be applied in
view of the constraint in $q_0$. To see what happens then,
consider the poles of the integrand of \Eqref{sif} in
the complex $q_0$ plane. 

There are four poles (two in the upper
and two in the lower half-plane) coming from the propagators in
the two-particle Green function:
$$ q_0 = \pm \half P_0 - \sqrt{m^2 + (\half \vec{P} \mp \vec{q})^2} +i\veps,
\,\,\, {\rm and} \,\,\,
 q_0 = \pm \half P_0 + \sqrt{m^2 + (\half \vec{P} \mp \vec{q})^2} -i\veps. $$
We can see that a simultaneous sign reflection of $P_0$ and $q_0$
interchanges the
poles of the upper half-plane with the poles
of the lower half-plane. The same symmetry exists for the
singularities of $\Phi$. Therefore, in order for $\Si$ to be 
even in $P_0$
the integration over $q_0$ must be independent of the choice
of the half-plane where we close the contour. 
In performing a 3-D reduction,
however, one usually neglects the contribution of certain
poles, hence the result becomes dependent on the contour.
In that case $\Si$ is not anymore an even function of $P_0$,
consequently it cannot be a function of $P^2$ only.
In this sense the standard Lorentz structure of the self-energy is violated.
(An example of the QP prescription, which is covariant under the  
Lorentz group but violates the charge conjugation symmetry,
is the spectator approximation of Gross \cite{Grs69}.
In this approximation, one of the particles inside the loop is restricted
to its mass shell, therefore only a single pole is taken in calculating
the $q_0$ integral.)

Similar arguments apply for the spin-1/2 particle
self-energies. Consider the dressed fermion propagator given by
\beq
\eqlab{dprop}
S(\sla{P}) = \left[ \sla{P} - m - \Si (\sla{P}) +i\eps \right]^{-1},
\eeq
where $\Si (\sla{P})$ is the self-energy.
For simplicity we work in the  c.m.\ frame, where $P=(P_0,\,\vec{0})$.
In this frame the Dirac structure of the self-energy can 
be represented as
\beq
\eqlab{self}
\Si(P_0)=\Si_+(P_0)\ga_+ + \Si_-(P_0) \ga_-,
\eeq
where $\gamma_\pm=\half (I \pm \ga_0).$
A similar decomposition holds for the propagator:
\beq
\eqlab{propcms}
S(P_0) = S^{(+)}(P_0)\ga_+ + S^{(-)}(P_0) \ga_-,
\eeq
with $S^{(\pm)}(P_0) = \pm [ P_0 \pm (- m - \Si_\pm (P_0)
+i \epsilon)]^{-1}$.
Obviously, $S^{(+)}$ corresponds to the positive and $S^{(-)}$ to the
negative energy-state propagation.

It is easy to see that, if the self-energy can be written in the
general covariant form \eref{form}, then
the following identity holds,
\beq
\eqlab{posneg}
\Si_{r} (P_0) = \Si_{-r} (-P_0),\,\,\,\, r=\pm 1,
\eeq
and vise versa (in the c.m.\ frame).
This identity is particularly useful to test numerically 
\Eqref{form} in models based on QP equations which are usually
solved for partial waves in the c.m.\ system.

Performing the standard renormalization procedure by subtracting
the counter-term: $Z_2 (m_0-m)+(1-Z_2)(\sla{P}-m)$,
where $m_0$ is the bare mass, and $Z_2$ is the field
renormalization constant,
we find that the on-shell renormalization scheme requires
\bea
Z_2 (m_0 - m)&=& \Si_+ (m) =\Si_- (-m) , \no\\ 
1-Z_2&=& 
\left.\frac{\pa \Si_+ (P_0)}{\pa P_0}\right|_{P_0 = m}=
-\left.\frac{\pa \Si_- (P_0)}{\pa P_0}\right|_{P_0 = -m} .
\eea
Obviously, it is not possible to satisfy these relations if 
\Eqref{posneg} is violated. In other words, the violation of
the extended Lorentz symmetry leads to the different renormalization
of the positive and negative energy states. This can be understood,
as the violation of the charge conjugation symmetry in a 
Lorentz-covariant framework implies 
violation of CPT symmetry.

To recapitulate, relativistic equations obtained from the BS equation
via the 3-D reduction which discriminates between the 
positive and negative energy poles 
(\eg by putting particles on-shell, or using the positive energy
projection operators) lead to results which do not have the
standard Lorentz structure, even if the symmetry under the
full Lorentz group remains intact. Such equations necessarily
violate the charge conjugation and CPT symmetries, and thus lead to the
breakdown of the usual renormalization procedures which rely on
constructing the counter-terms from a CPT invariant Lagrangian.
Also, one then cannot use the standard
covariant arguments to construct the transformation properties of 
the calculated amplitudes (as well as any other functions
involving loop corrections),
which for instance is needed to incorporate
the basic interaction in more particle systems.

\section{Manifestly covariant three-dimensional equation}
One of the ways to perform a 3-D reduction 
consistent with charge conjugation
and unitarity is by removing the poles
of the interaction in $q_0$ complex plane, while  
treating  exactly the poles of the two-particle propagator.
This procedure is realized in the equal-time (ET) 
approximation \cite{MaW87,tlh}. 
In this approximation the poles are removed from the interaction
piece by fixing the relative-energy variable $q_0$ in some way. 
Most frequently the constraint $q_0=0$, or its 
Lorentz-invariant generalization, $P\cdot q = 0$, is used. 
Moreover, the two-particle propagator is sometimes modified
to include approximately the crossed graphs \cite{MaW87,TiT94}.

On the other hand, it is well known that the $P\cdot q = 0$ constraint
is troublesome in the inelastic or more-particle problems, see, e.g.,
the introductory remarks in Refs.~\cite{DiW93,PhW98}.  
The weak point resides in the fact that the constraint
is embedded through a $\de$-function. In the following
we formulate a 3-D formulation
which exhibits manifest Lorentz covariance, does not make use of
$\de$-functions, and for the elastic two-body problem is equivalent to
the ET approximation.

Recall the two-particle Bethe-Salpeter equation in the momentum-space: 
\beq
 T(p',p)=V(p',p)+ i\int\! \frac{\dd^4 q}{(2\pi)^4}\,V(p',q)\,G(q)
\,T(q,p),
\eeq
where we assume $p',\,p$ and $q$ are the relative four-momenta of the
final, initial and intermediate state, respectively.
To transit to the 3-D formulation we impose the condition
that the interaction is insensitive to the off-shellness along
the direction defined by unit four-vector $n_\mu$. For the two-body
case this means that $V$ and $T$ entering the scattering equation
 depend on the
projections of the relative four-vectors onto a 3-D hyperplane
orthogonal to $n_\mu$. Defining the projection operator:
$O_{\mu\nu}=\rg_{\mu\nu}-n_\mu n_\nu$, 
we write the corresponding equation as follows:
\beq
\eqlab{compact}
T(\tp',\tp)=V(\tp',\tp)+ i\int\,\frac{\dd^4 q}{(2\pi)^4} \,
 V(\tp',\tq)\, G(q)\, T(\tq,\tp),
\eeq
where $\tp_{\mu} = O_{\mu\nu} p^{\nu}$,   
and similarly for $\tp',\,\tq$. 
Equation \eref{compact} is manifestly covariant,
and on the other hand it can easily be reduced to 
the 3-D form. 
For example, let us choose the frame where 
$n=(1,0,0,0)$, and therefore $V$ and $T$ are independent
of the 0-th component of relative momenta (since any scalar product
will depend only on the spatial components, e.g., $\tq\cdot\ga=-
\vec{q}\cdot\vec{\ga}$). The integration over $q_0$ in \Eqref{compact}
can now be readily done leading to the 3-D equation. 

Obviously, the newly introduced  four-vector $n$ will enter the 
final covariant forms. To prevent this dependence one may choose 
it along some physical four-momentum, for instance the
total momentum of the system, \ie 
\beq
n_\mu=P_\mu/\sqrt{P^2}.
\eeq
 It is then easy to see
that for the two-body elastic
scattering \Eqref{compact} in the c.m.\ system becomes
equivalent to the usual ET approximation.

\section{Box graphs and the one-body limit}
Analyzing the box and the crossed-box graphs in QFT,
for a neutral particle exchange, Gross
revealed a cancellation among various pole 
contributions, and proved that the only pole which survives in
the limit is that of the heavy particle in the intermediate state of 
the box graph \cite{Grs69,Grs94}. 
This led him to formulate the spectator equation where the heavy particle 
is on the mass-shell. Obviously, the heavier is the spectator, the
closer should the Gross (spectator) equation be to the QFT result.
Therefore, for instance the \pn
system would seem to be a particularly good application
for this equation, since the nucleon is much heavier than the pion. 
Recently, however, Gross and Surya, applying
the spectator equation to $\pi N$ system, have argued that the 
light particle (the pion) must be taken as the spectator
\cite[Sec.\ II.A]{GrS93}. 
Studying the box and the crossed-box graphs at threshold, they have
conjectured that ``the essential difference is the mass of the
exchanged particle''. 
We have examined their conjecture, studying the graphs for more
general situations, and find that the argument should be related more
to the type of the OPE potential, rather than
to the mass of the exchanged particle. 

\begin{figure}[t]
\begin{center}
\epsffile{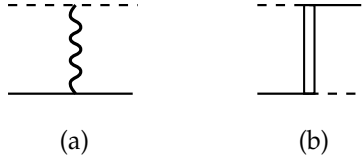}
\vspace{-0.2cm}
\end{center}
\caption[F1]{The $t$-channel (a) and $u$-channel (b) exchange potentials.}
\figlab{boxpotf}
\end{figure}

Namely, we consider two types of the potentials, see \Figref{boxpotf}:
(a) $t$-exchange potential, and 
(b) $u$-exchange potential.
(We shall refer to the dashed line particle as to pion and the solid 
line as to nucleon with corresponding masses $m_\pi$ and $m_N$, 
the exchange particle mass is denoted as $\mu$.) 
\begin{figure}[h]
\begin{center}
\vspace{-0.5cm}
\epsffile{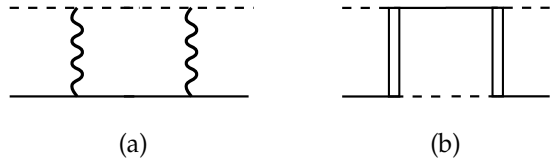}
\end{center}
\vspace{-0.7cm}
\caption[F3]{The box graphs obtained by iterating once the
potentials of \Figref{boxpotf}.}
\figlab{boxboxf}
\end{figure}
Substituting these potentials
into the scattering equation, \Figref{bsef}, and
iterating once, we obtain the box graphs
 depicted in \Figref{boxboxf} (a)
and (b), respectively. 
Note that in QFT, due to the crossing symmetry,  one 
in addition has the corresponding crossed-box graphs. 

\begin{figure}[t]
\begin{center}
\epsffile{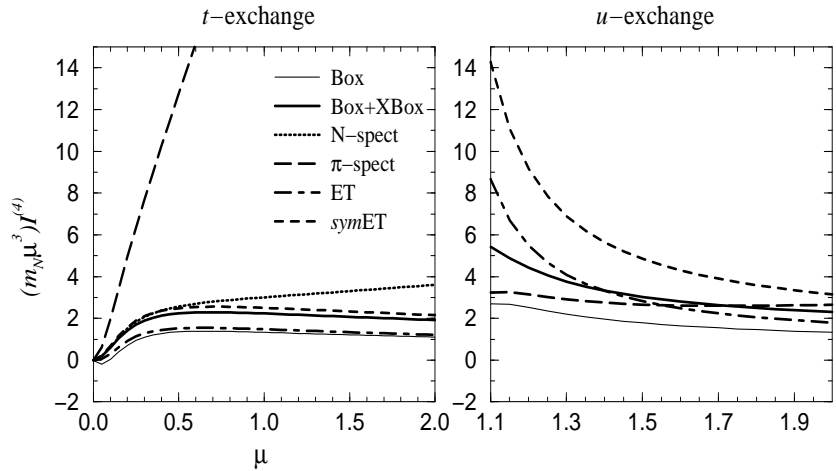}
\end{center}
\caption[F4]{Results for $m_N=1,\,m_\pi=0.01,\, \sqrt{s}=1.1$ 
as the function of the exchange particle mass. }
\figlab{ex1}
\end{figure}

We have calculated such box and crossed-box graphs 
in 4-D field theory numerically
and compared with the box graph calculation within various
quasipotential formulations. Namely, the nucleon 
and the pion {\it spectator} \cite{Grs69,Grs94,GrS93}, the 
{\it equal-time} \cite{tlh}, 
and the {\it symmetrized equal-time} \cite{MaW87,TiT94}.

The results of these calculations for the case close to the one body limit
(the nucleon is much heavier that the pion) is plotted in \Figref{ex1},
as a function of the exchange particle mass.\footnote{Note that we
multiply the results by $\mu^3 m_N$  in order to obtain
reasonable values for various limiting cases.}
The energy is fixed slightly above the threshold, $\sqrt{s}=1.1 m_N$, 
and $t=0$. One can see that for the $t$-exchange potential 
the one-body limit is achieved
in the symmetrized ET formulation independently of the
mass of the exchanged particle. The nucleon spectator indeed
deviates from the limit for large $\mu$, however the pion spectator
does not produce a better result in this situation.

On the other hand,  in 
the $u$-exchange case, both the nucleon spectator and symmetrized
ET disagree substantially
 with the QFT result (the spectator calculation is an order
of magnitude larger and hence beyond the scale of the figure). 
The pion spectator is in a much better agreement.
Thus, we conclude that the difference
between the $NN$ and $\pi N$ situation encountered
by Gross and Surya \cite{GrS93} appears
due to the different {\it type of the potential}.

It would be interesting to see if there is a possibility to develop
a prescription which would give the proper limit in both the
$t$- and $u$-exchange cases.  It should be emphasized though,
that the one-body limit situation is physically very different
for the two cases: for the $t$-exchange potential it corresponds to
the light particle moving in an external potential of the heavy particle,
while in the  $u$-exchange case the heavy particle obviously does not act
as a static external source, and therefore there seems to be
no correspondence to any one-body situation.

\section{$\pi N$ scattering}

We have studied the ET approximation of the BS equation in a dynamical 
model for $\pi N$ scattering. The corresponding equation, \Figref{bsef},
is solved for the $\pi N$ partial-wave amplitudes with the OPE potential
represented by $N$(938), $N^*$(1450), $\De$(1232), D$_{13}$(1525), 
S$_{11}$(1555), $\rho$(770)
and $\sigma$(550) exchanges, see \Figref{potf}.
The model is very close to the one presented earlier \cite{PT97},
even though presently
we have used a different form of the $\pi N\De$ coupling \cite{Pas98},
and D$_{13}$ and S$_{11}$ exchanges are included in addition. The latter
has a considerable effect only in the S$_{11}$ partial wave.

The model parameters (coupling constants, resonance and cutoff masses)
 were adjusted to reproduce the
low-energy quantities, such as scattering lengths, volumes 
and ranges, and the
energy behavior of the phase-shifts. The resulting description of the
phase-shifts up to 600 MeV pion kinetic lab energy is depicted in 
\Figref{pin600f}.

\begin{figure}[t]
\epsffile{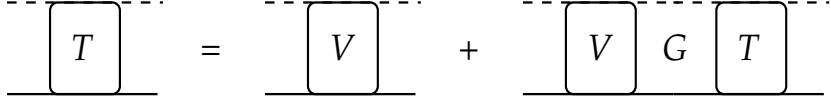}
\caption[F2]{Diagrammatic form of a relativistic 
two-body scattering equation.}
\figlab{bsef}
\vspace{-0.5cm}
\end{figure}
\begin{figure}[h]
\epsffile{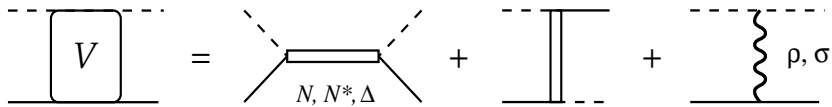}
\caption{The tree-level $\pi N$  potential, the driving force of the
scattering equation.}
\figlab{potf}
\vspace{-0.5cm}
\end{figure}

\epsfxsize=16.0cm
\begin{figure}[t]
\begin{center}
\epsffile{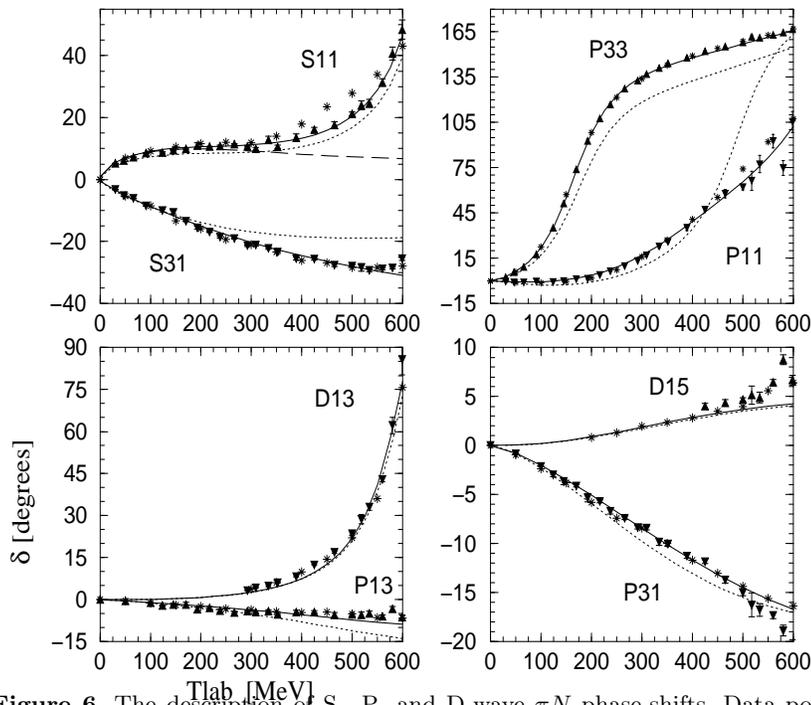}
\vspace{-0.8cm}
\caption{The description of S-, P- and D-wave $\pi N$ phase-shifts.
  Data points are from the SM95 \cite{ASW95}
 (triangles) and KH80 \cite{KH80} (stars)
partial-wave analyses. Solid lines represent the model solution. 
Dotted lines represent the calculation where the principal value
part of the rescattering integrals is switched off
(\ie the K-matrix approximation with the same set of
parameters). Dashed line for the S$_{11}$ shows the calculation
when the pole contribution of the S$_{11}$ resonance is switched off.}
\figlab{pin600f}
\end{center}
\vspace{-1cm}
\end{figure}

\section{Conclusions}
The relativistic scattering and bound state problems are often formulated
in terms of a 3-D (or quasipotential) relativistic equation of 
the Lippmann-Schwinger type. Such equations can be obtained from
the manifestly covariant (3+1)-dimensional Bethe-Salpeter equation by
integrating out the time variable in some approximate way,
thus performing the so called 3-D reduction.
Adopting the 3-D formulation in favor of the 
4-D one leads to major technical simplifications, 
since the field-theoretical BS kernel 
may, in principle, contain many singularities in the time variable.
However, the charge conjugation symmetry can
easily be violated in performing such a  reduction.
On the other hand, it plays
an important role in obtaining the standard Lorentz
structure of the loop corrections. Therefore, the
equations which respect charge conjugation symmetry
are preferable, and thus the choice among the 
infinite number of possible relativistic 3-D equations is somewhat 
restricted in this way. 

We have presented a 3-D reduction which is manifestly
covariant under the complete set of Lorentz transformations
as well as charge conjugation. The two-body equation, obtained
by using this reduction, in the c.m.\ system is equivalent
to the Salpeter equation. 

We have studied  the one-body
limit for several 3-D equations with the OPE potential. 
In the limit a large qualitative difference 
is observed between the situation when the potential in question has
the form of $t$- or $u$-channel exchange. The 3-D equations,
such as the nucleon spectator \cite{Grs69,Grs94}
and the symmetrized ET \cite{MaW87,TiT94},
developed to satisfy the one-body limit for the $t$-type exchange potential,
have a poor agreement with the exact calculation if the
$u$-type exchange potential is used. 

The pion spectator 
approximation describes the $u$-exchange case better, but fails
in the other case. Therefore, in the situation where both
types of the potential are present, either of the spectator
equations cannot be justified. Analyzing the $\pi N$ situation with realistic
parameters we find that the ET type of prescriptions
can be fairly close to the QFT answer for both types of the potential,
and, hopefully, is a reasonable dynamical framework in this case.
We have therefore applied the ET approximation of the BS equation to
the description of the $\pi N$ scattering.

\section*{Acknowledgments}
We thank Dr.\ F.\ Coester for illuminating discussions on Lorentz
covariance. One of us (V.P.) had the pleasure to discuss
with Professor V.\ Mandelzweig during the conference.

\end{document}